\documentclass{llncs}

\usepackage{diagrams}

\def\fa{\forall}
\def\ex{\exists}
\def\ra{\rightarrow}
\def\la{\leftarrow}

\newbox\tempa
\newbox\tempb
\newdimen\tempc
\def\mud#1{\hfil $\displaystyle{\mathstrut #1}$\hfil}
\def\rig#1{\hfil $\displaystyle{#1}$}
\def\irulehelp#1#2#3{\setbox\tempa=\hbox{$\displaystyle{\mathstrut #2}$}%
                        \setbox\tempb=\vbox{\halign{##\cr
        \mud{#1}\cr
        \noalign{\vskip\the\lineskip}%
        \noalign{\hrule height 0pt}%
        \rig{\vbox to 0pt{\vss\hbox to 0pt{${\; #3}$\hss}\vss}}\cr
        \noalign{\hrule}%
        \noalign{\vskip\the\lineskip}%
        \mud{\copy\tempa}\cr}}%
                      \tempc=\wd\tempb
                      \advance\tempc by \wd\tempa
                      \divide\tempc by 2 }
\def\irule#1#2#3{{\irulehelp{#1}{#2}{#3}%
                     \hbox to \wd\tempa{\hss \box\tempb \hss}}}

\begin{document}
\title{Confluence as a cut elimination property}
\author{Gilles Dowek}
\date{}
\institute{
\'Ecole polytechnique and INRIA\\
LIX, \'Ecole polytechnique, 
91128 Palaiseau Cedex, France. \\
{\tt Gilles.Dowek@polytechnique.fr, http://www.lix.polytechnique.fr/\~{}dowek}}
\maketitle

\begin{abstract}
The goal of this note is to compare two notions, one coming from the
theory of rewrite systems and the other from proof theory: confluence
and cut elimination.  We show that to each rewrite system on terms, we
can associate a logical system: {\em asymmetric deduction modulo this
rewrite system} and that the confluence property of the rewrite system
is equivalent to the cut elimination property of the associated
logical system.  This equivalence, however, does not extend to rewrite
systems directly rewriting atomic propositions.
\end{abstract}

The goal of this note is to compare two notions, one coming from the
theory of rewrite systems and the other from proof theory: confluence
and cut elimination.

The confluence a rewrite system permits
to reduce the search space when we want to establish that
two terms are convertible. Similarly, the cut elimination property
of a logical system permits to reduce the search space, when we want to
establish that some proposition is provable.

Moreover, both properties can be used to prove the decidability
of convertibility or provability, when this reduction yields a finite
search space. Finally, both properties can be used to prove
independence results (i.e. that two terms are not convertible or that
a proposition is not provable), and in particular consistency results,
when this reduction yields an empty search space.

The goal of this note is to show that this similarity between
confluence and cut elimination can be seen as a consequence of the
fact that to each rewrite system ${\cal R}$ rewriting terms, we can
associate a logical system: {\em asymmetric deduction modulo ${\cal
    R}$}, a variant of deduction modulo introduced in \cite{DHK}, and
that the confluence property of the rewrite system is equivalent to
the cut elimination property of the associated logical system.  More
precisely, we establish a parallel between
\begin{itemize}
\item an equality $t = u$ and a sequent $P(t) \vdash P(u)$, 
\item the notion of conversion sequence and that of proof,
\item the notion of peak and that of cut, and
\item the notion of valley sequence and that of cut free proof.
\end{itemize}

Both valley sequences and cut free proofs may be called {\em analytic}
as they exploit the information present in their conclusion and its
sub-parts but no other information. 

Finally, we relate a method used to prove cut elimination by defining
an algorithm transforming proofs step by step until all cuts are
removed (see, for instance, \cite{Giraflor}) and a method used to
prove confluence by defining an algorithm transforming rewrite
sequences step by step until all peaks are removed (see, for instance,
\cite{ChurchRosser,Newman,BachmairDershowitz}). As an example, we
reformulate Newman's confluence theorem \cite{Newman} as a cut
elimination theorem.

Asymmetric deduction modulo can be extended by allowing not only rules
rewriting terms in propositions, but also directly atomic propositions. With
such rules, confluence and cut elimination do not coincide anymore and
confluence is not a sufficient analyticity condition: it must be
replaced by cut elimination.

\section{Asymmetric deduction modulo}

In deduction modulo \cite{DHK}, the notions of language, term and
proposition are 
that of first-order predicate logic. But a theory is formed with a set of
axioms $\Gamma$ {\em and a congruence $\equiv$} defined on
propositions. 
Deduction rules are modified to take this congruence into account. For
instance, the right rule of conjunction is not stated as usual 
$$\irule{\Gamma \vdash A, \Delta~~~\Gamma \vdash B, \Delta}
        {\Gamma \vdash A \wedge B, \Delta}
        {}$$
as the conclusion need not be exactly $A \wedge B$ but may be
only convertible to this proposition, hence it is stated
$$\irule{\Gamma \vdash A, \Delta~~~\Gamma \vdash B, \Delta}
        {\Gamma \vdash C, \Delta}
        {\mbox{if $C \equiv A \wedge B$}}$$
All rules of sequent calculus, or natural deduction, may be defined in
a similar way. 

In this note, we consider only congruences defined by a rewrite system
on terms. A {\em rewrite rule} is a pair of terms $\langle l,
r\rangle$, written $l \ra r$, such that $l$ is not a variable. A {\em
  rewrite system} is a set of rules. Given such a system, the relation
$\ra^{1}$ is the smallest relation defined on terms and on
propositions compatible with the structure of terms and propositions
such that for all substitutions $\theta$ and all rewrite rules $l \ra
r$ of the rewrite system $\theta l \ra^{1} \theta r$. The relation
$\ra^{+}$ is the transitive closure of $\ra^{1}$, the relation
$\ra^{*}$ is its reflexive-transitive closure and the relation
$\equiv$ its reflexive-symmetric-transitive closure.  Notice that
rewriting does not change the logical structure of a proposition, in
particular an atomic proposition only rewrites to an atomic
proposition.

A {\em conversion sequence} is a finite sequence of terms or
propositions $C_{1}, ...,
C_{n}$ such that for each $i$ either $C_{i} \ra^{1} C_{i+1}$ or 
$C_{i} \la^{1} C_{i+1}$. Obviously two terms or two propositions $A$
and $B$ are convertible if there is a conversion sequence whose first
element is $A$ and last element is $B$. 
A {\em peak} in such a sequence is an index $i$ such that $C_{i-1}
\la^{1} C_{i} \ra^{1} C_{i+1}$. A sequence is called a {\em valley
sequence} if it contains no peak, i.e. if it has the form 
$A \ra^{1} ... \ra^{1} \la^{1} ... \la^{1} B$.

For example, in arithmetic, we can define a congruence with the
following rules
$$0 + y \rightarrow y$$
$$S(x) + y \rightarrow S(x+y)$$
$$0 \times y \rightarrow 0$$
$$S(x) \times y \rightarrow x \times y + y$$
In the theory formed with the axiom $\forall x~x = x$ and this congruence,
we can prove, in sequent calculus modulo, that the number $4$ is even
$$\irule{\irule{\irule{}
                      {4 = 4 \vdash 2 \times 2 = 4}
                      {\mbox{Axiom}}
               }
               {\fa x~x = x \vdash 2 \times 2 = 4}
               {\mbox{$(x,x = x,4)$ $\forall$-left}}
        }
        {\fa x~x = x \vdash  \exists x~2 \times x = 4}
        {\mbox{$(x,2 \times x = 4,2)$ $\exists$-right}}$$
The sequent $4 = 4 \vdash 2 \times 2 = 4$, that requires a tedious proof
in usual formulations of arithmetic, can simply be proved with the axiom
rule here, as $(4 = 4) \equiv (2 \times 2 = 4)$.

Deduction modulo a congruence defined by a rewrite system on terms
uses this rewrite system only through the congruence it generates. The
way these congruences are established is not taken into consideration.
Thus, cut free proofs are analytic in the sense that they do not use
the cut rule but not in the sense they establish congruences with
valley sequences. Hence, we introduce a weaker
system, {\em asymmetric deduction modulo}, where
propositions can only be reduced. The rules of 
asymmetric sequent calculus modulo are given in figure \ref{SeqMod}.

\begin{figure}
\noindent\framebox{\parbox{\textwidth
}{
{\small
$$
\begin{array}{c}
\irule{}
        {\Gamma, A_{1} \vdash A_{2}, \Delta}
        {\mbox{$(A)$ Axiom if $A_{1} \ra^{*} A \la^{*} A_{2}$}}\\
\irule{\Gamma \vdash C_{1}, \Delta ~~~ \Gamma, C_{2} \vdash \Delta}
        {\Gamma \vdash \Delta}
        {\mbox{$(C)$ Cut if $C_{1} \la^{*} C \ra^{*} C_{2}$}}\\
\irule{\Gamma, A_1, A_2 \vdash \Delta}
        {\Gamma, A \vdash \Delta}
        {\mbox{contr-left if $A_{1} \la^{*} A \ra^{*} A_2$}}\\
\irule{\Gamma \vdash A_1,A_2,\Delta}
        {\Gamma \vdash A,\Delta}
        {\mbox{contr-right if $A_{1} \la^{*} A \ra^{*} A_2$}}\\
\irule{\Gamma \vdash \Delta}
        {\Gamma, A \vdash \Delta}
        {\mbox{weak-left}}\\
\irule{\Gamma \vdash\Delta}
        {\Gamma \vdash A,\Delta}
        {\mbox{weak-right}}\\
\irule{\Gamma \vdash A, \Delta ~~~ \Gamma, B \vdash \Delta}
        {\Gamma, C \vdash  \Delta}
        {\mbox{$\Rightarrow$-left if $C \ra^{*} (A \Rightarrow B)$}}\\
\irule{\Gamma, A \vdash B, \Delta}
        {\Gamma \vdash  C, \Delta}
        {\mbox{$\Rightarrow$-right if $C \ra^{*} (A \Rightarrow B)$}}\\
\irule{\Gamma, A, B \vdash \Delta}
        {\Gamma, C \vdash  \Delta}
        {\mbox{$\wedge$-left if $C \ra^{*} (A \wedge B)$}}\\
\irule{\Gamma \vdash A, \Delta~~~\Gamma \vdash B, \Delta}
        {\Gamma \vdash C, \Delta}
        {\mbox{$\wedge$-right if $C \ra^{*} (A \wedge B)$}}\\
\irule{\Gamma, A \vdash \Delta ~~~ \Gamma, B \vdash \Delta}
        {\Gamma, C \vdash  \Delta}
        {\mbox{$\vee$-left if $C \ra^{*} (A \vee B)$}}\\
\irule{\Gamma \vdash A, B, \Delta}
        {\Gamma \vdash C, \Delta}
        {\mbox{$\vee$-right if $C \ra^{*} (A \vee B)$}}\\
\irule{}
        {\Gamma, A \vdash \Delta}
        {\mbox{$\bot$-left if $A \ra^{*} \bot$}}\\
\irule{\Gamma, [t/x]A \vdash \Delta}
        {\Gamma, B \vdash \Delta}
        {\mbox{$(x,A,t)$~$\fa$-left if $B \ra^{*} \fa x~A$}}\\
\irule{\Gamma \vdash A, \Delta}
        {\Gamma \vdash B, \Delta}
        {\mbox{$(x,A)$~$\fa$-right if $B \ra^{*} \fa x~A$ and
         $x \not\in FV(\Gamma \Delta)$}}\\
\irule{\Gamma, A \vdash \Delta}
        {\Gamma, B \vdash \Delta}
        {\mbox{$(x,A)$~$\ex$-left if $B \ra^{*} \ex x~A$ and 
         $x \not\in FV(\Gamma \Delta)$}}\\
\irule{\Gamma \vdash [t/x]A, \Delta}
        {\Gamma \vdash B, \Delta}
        {\mbox{$(x,A,t)$~$\ex$-right if $B \ra^{*} \ex x~A$}}
\end{array}~~~~~~~~~~~~~~~~~~~~~~~~~~~~~~~~~~~~~~$$
\caption{Asymmetric sequent calculus modulo}
\label{SeqMod}
}}}
\end{figure}
\renewcommand{\arraystretch}{1}

\section{Cut elimination in atomic symmetric deduction modulo}

We first consider a fragment of symmetric deduction modulo where all
propositions are atomic. This system contains only the axiom rule, the
cut rule and the structural rules (weakening and contraction). 

To relate proofs in atomic deduction modulo and rewrite sequences, we
prove that the sequent $P(t) \vdash P(u)$ is provable in atomic
deduction modulo if and only if $t \equiv u$. This is a direct
consequence of the following proposition.
 
\begin{proposition}
\label{convertible}
In atomic deduction modulo, 
the sequent $\Gamma \vdash \Delta$ is provable 
if and only if $\Gamma$ contains a proposition $A$ and $\Delta$
a proposition $B$ such that $A \equiv B$.
\end{proposition}

\proof{If $\Gamma$ contains a proposition $A$ and $\Delta$
a proposition $B$ such that $A \equiv B$, then the sequent $\Gamma
\vdash \Delta$ can be proved with the axiom rule. 

Conversely, we prove by induction over proof structure that
if the sequent $\Gamma \vdash \Delta$ is provable, then 
$\Gamma$ contains a proposition $A$ and $\Delta$ a proposition $B$
such that $A \equiv B$. 

This is obvious 
if the last rule of the proof is the axiom rule.
If the last rule is a structural rule, we simply apply the induction
hypothesis.
Finally, if the last rule is the cut rule, then the proof has the form 
$$\irule{\irule{\pi_{1}}
               {\Gamma \vdash C_{1}, \Delta}
               {}
          ~~~
         \irule{\pi_{2}}
               {\Gamma, C_{2} \vdash \Delta}
               {}
        }
        {\Gamma \vdash \Delta}
        {\mbox{Cut}}$$
With $C_{1} \equiv C_{2}$. By induction hypothesis
\begin{itemize}
\item $\Gamma$ and $\Delta$ contain two convertible propositions or
$\Gamma$ contains a proposition convertible to $C_{1}$, 
\item 
$\Gamma$ and $\Delta$ contain two convertible propositions or 
$\Delta$ contains a proposition convertible to $C_{2}$. 
\end{itemize}
Thus, in all cases, $\Gamma$ and $\Delta$ contain two convertible propositions.
} 

\medskip

The next proposition shows that atomic deduction modulo has the cut
elimination property, i.e. that all provable sequents have a cut free
proof. It is known, in general, that deduction modulo a congruence
defined by a rewrite system on terms has the cut elimination property
\cite{DowekWerner-normalization-98}, but for the case of atomic
deduction modulo, this is a direct consequence of proposition
\ref{convertible}.

\begin{proposition}
In atomic deduction modulo, all provable sequents have a cut free proof.
\end{proposition}

\proof{If a sequent $\Gamma \vdash \Delta$ is provable then, by
proposition \ref{convertible}, $\Gamma$ and $\Delta$ contain two
convertible propositions and thus the sequent $\Gamma \vdash \Delta$
can be proved with the axiom rule.}

\medskip

We can also define a proof reduction algorithm that reduces cuts step
by step.

\begin{definition}[Proof reduction]
\label{redsym}

Consider a proof of the form
$$\irule{\irule{\pi_{1}}{\Gamma \vdash C_{1}, \Delta}{}
         ~~~~~~~~~~~~~~
         \irule{\pi_{2}}{\Gamma, C_{2} \vdash \Delta}{}
        }
        {\Gamma \vdash \Delta}
        {\mbox{Cut}}$$
where $\pi_{1}$ and $\pi_{2}$ are cut free proofs.

The multisets $\Gamma$ and $C_{1}, \Delta$ contain two convertible
propositions. Similarly, $\Gamma, C_{2}$ and $\Delta$ contain two
convertible propositions.  Thus,
\begin{itemize}
\item $\Gamma$ and $\Delta$ contain two convertible propositions or
$\Gamma$ contains a proposition convertible to $C_{1}$, 
\item 
$\Gamma$ and $\Delta$ contain two convertible propositions or 
$\Delta$ contains a proposition convertible to $C_{2}$. 
\end{itemize}
Thus, as $C_{1} \equiv C_{2}$, $\Gamma$ and $\Delta$ contain two convertible
propositions in all cases and this proof reduces to 
$$\irule{} {\Gamma \vdash \Delta} {\mbox{Axiom}}$$ 

When a proof contains a cut, the proofs of the premises of the highest 
cut are obviously cut free and this proof reduction algorithm applies.
It terminates because it removes a cut at each step. 
Thus, it produces a cut free proof after a finite number of steps.
\end{definition}
 
\section{Cut elimination in atomic asymmetric deduction modulo}

Let us now turn to asymmetric deduction modulo, still in the atomic
case.  We prove that in asymmetric atomic deduction modulo, a sequent
$P(t) \vdash P(u)$ is provable if and only if $t \equiv u$ and that
this sequent has a cut free proof if and only if $t$ and $u$ have a
common reduct. Thus, proofs in asymmetric deduction modulo correspond
to rewrite sequences and cut free proofs to valley sequences.

\begin{proposition}
\label{withcuts}
In asymmetric atomic deduction modulo, the sequent $\Gamma \vdash \Delta$ is
provable if and only if $\Gamma$ contains a proposition $A$ and $\Delta$
a proposition $B$ such that $A \equiv B$.
\end{proposition}

\proof{Obviously, if the sequent $\Gamma \vdash \Delta$ is provable in 
asymmetric deduction modulo, then it is provable in symmetric deduction
modulo and $\Gamma$ and $\Delta$ contain two convertible propositions.

The converse is slightly more difficult than in the symmetric case
because the axiom rule does not apply directly.
Assume there are propositions $A$ in $\Gamma$ and $B$ in $\Delta$
such that $A \equiv B$. 
Then there is a rewrite sequence $A = C_{1}, ..., C_{n} = B$ 
joining $A$ and $B$. We prove, by induction on the number of peaks in
this sequence that the sequent $\Gamma \vdash \Delta$ is provable.  

If the sequence contains no peak, then $A$ and $B$ have a common
reduct and the sequent 
$\Gamma \vdash \Delta$ can be proved with the axiom rule.
Otherwise, there is a peak $i$ in this sequence.
The sequences $C_{1}, ..., C_{i}$ and $C_{i}, ..., C_{n}$ contain
fewer peaks than $C_{1}, ..., C_{n}$, thus, by induction
hypothesis, the sequents $\Gamma \vdash C_{i}, \Delta$ and 
$\Gamma, C_{i} \vdash \Delta$ have proofs $\pi_{1}$ and $\pi_{2}$.
We build the proof
$$\irule{\irule{\pi_{1}}
               {\Gamma \vdash C_{i}, \Delta}
               {}
         ~~~
        \irule{\pi_{2}}
              {\Gamma, C_{i} \vdash \Delta}
              {}
        }
        {\Gamma \vdash \Delta}
        {\mbox{$(C_{i})$~Cut}}$$
}

\begin{proposition}
\label{cutfree}
In asymmetric deduction modulo, the sequent $\Gamma \vdash \Delta$ has
a cut free proof if and only $\Gamma$ contains a proposition $A$ and
$\Delta$ a proposition $B$ such that $A$ and $B$ have a common reduct.
\end{proposition}

\proof{
If $\Gamma$ contains a proposition $A$ and
$\Delta$ a proposition $B$ such that $A$ and $B$ have a common reduct, then 
the sequent $\Gamma \vdash \Delta$ can be proved with the axiom rule.
Conversely, if the sequent $\Gamma \vdash \Delta$ has a cut free proof, 
we prove, by induction over proof structure, that 
$\Gamma$ contains a proposition $A$ and $\Delta$ a proposition $B$ 
such that $A$ and $B$ have a common reduct. 
This is obvious if the last rule is an axiom rule. If the last rule is 
a structural rule, we apply the induction hypothesis.}

\medskip

We can now state our main proposition that relates confluence and cut
elimination.

\begin{proposition}[Main proposition]
\label{main}
The cut rule is redundant in asymmetric atomic deduction modulo a
rewrite system if and only if this rewrite system is confluent. 
\end{proposition}

\proof{Assume that the rewrite system is confluent.
If the sequent $\Gamma \vdash \Delta$ is provable then,
by proposition \ref{withcuts},
$\Gamma$ and $\Delta$ contain two convertible propositions.
By confluence, they have a common reduct and thus,
by proposition \ref{cutfree},
the sequent $\Gamma \vdash \Delta$ has a cut free proof.

Conversely, assume that the cut rule is redundant. If $A \equiv B$
(resp. $t \equiv u$) then 
by proposition \ref{withcuts},
the sequent $A \vdash B$ (resp. $P(t) \vdash
P(u)$) is provable, thus
it has a cut free proof and,
by proposition \ref{cutfree},
$A$ and $B$
(resp. $t$ and $u$) have a common reduct.} 

\section{Proof reduction}

Like in the symmetric case (definition \ref{redsym}) we want to
design an algorithm reducing proofs and eliminating cuts step by step.
In the asymmetric case however this algorithm is a little more
involved as it needs to perform shorter steps to reconstruct reductions 
instead of conversions. This algorithm is a reformulation in sequent calculus
of Newman's algorithm that reduces rewrite sequences eliminating peaks,
step by step. To define it, we need to use the local confluence 
of the rewrite system and, to prove its
termination, the termination of the system.

\begin{definition}[Proof reduction]
\label{asymmetricreduction}
Consider a proof of the form
$$\irule{\irule{\pi_{1}}{\Gamma \vdash C_{1}, \Delta}{}
         ~~~~~~~~~~~~~~
         \irule{\pi_{2}}{\Gamma, C_{2} \vdash \Delta}{}
        }
        {\Gamma \vdash \Delta}
        {\mbox{$(C)$ Cut}}$$
where $\pi_{1}$ and $\pi_{2}$ are cut free proofs.

As $\pi_{1}$ is cut free, $\Gamma, C_{1}$ and $\Delta$ contain 
two propositions that have a common reduct.
Similarly, $\Gamma$ and $C_{2}, \Delta$ contain 
two propositions that have a common reduct.

If $\Gamma$ and $\Delta$ contain two propositions that have a
common reduct $C'$, then this proof reduces to
$$\irule{} {\Gamma \vdash \Delta} {\mbox{$(C')$ Axiom}}$$ 
Otherwise, $\Gamma$ contains a
proposition $A$ that has a common reduct $B$ with $C_{1}$ and $\Delta$
contains a proposition $E$ that has a common reduct $D$ with
$C_{2}$. Let us write $\Gamma = \Gamma', A$ and $\Delta = E, \Delta'$.
We have 
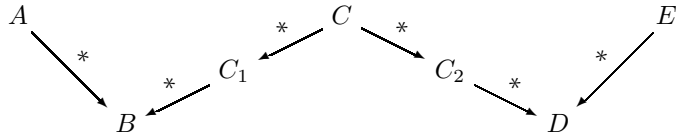
\begin{diagram}[height=1em,width=2em]
A&&&&&&C&&&&&&E\\
&\rdTo(2,4)^*&&&&\ldTo^*&&\rdTo^*&&&&\ldTo(2,4)^*&\\
&&&&C_1&&&&C_2&&&&\\
&&&\ldTo^*&&&&&&\rdTo^*&&&\\
&&B&&&&&&&&D&&\\
\end{diagram}

If $B = C$ or $C = D$ then either $D$ or $B$ is a common reduct of $A$
and $E$ and this proof reduces to 
$$\irule{}
        {\Gamma \vdash \Delta}
        {\mbox{$(D)$ Axiom}}$$
or to
$$\irule{}
        {\Gamma \vdash \Delta}
        {\mbox{$(B)$ Axiom}}$$
Otherwise we have  
$B \la^{+} C \ra^{+} D$
and there are propositions $C'_{1}$ and $C'_{2}$ such that 
%$B \la^{*} C'_{1} \la^{1} C \ra^{1} C'_{2} \ra^{*} D$.
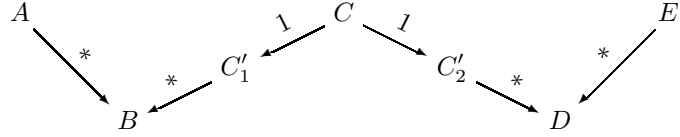
\begin{diagram}[height=1em,width=2em]
A&&&&&&C&&&&&&E\\
&\rdTo(2,4)^*&&&&\ldTo^1&&\rdTo^1&&&&\ldTo(2,4)^*&\\
&&&&C'_1&&&&C'_2&&&&\\
&&&\ldTo^*&&&&&&\rdTo^*&&&\\
&&B&&&&&&&&D&&\\
\end{diagram}
If there is a proposition $C'$ such that 
$C'_{1} \ra^{*} C' \la^{*} C'_{2}$, then we have 
%$A \ra^{*} B \la^{*} C'_{1} \ra^{*} C' \la^{*} C'_{2} \ra^{*} D \la^{*} E$ 
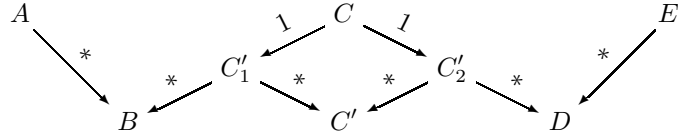
\begin{diagram}[height=1em,width=2em]
A&&&&&&C&&&&&&E\\
&\rdTo(2,4)^*&&&&\ldTo^1&&\rdTo^1&&&&\ldTo(2,4)^*&\\
&&&&C'_1&&&&C'_2&&&&\\
&&&\ldTo^*&&\rdTo^{*}&&\ldTo^{*}&&\rdTo^*&&&\\
&&B&&&&C'&&&&D&&\\
\end{diagram}
and this proof reduces to

\hspace*{5.1cm}{\small $\irule{\irule{\irule{}
                      {\Gamma', A \vdash C'_{1}, C'_{2}, E, \Delta'}
                      {\mbox{$(B)$ Axiom}}
                ~~~~~~~~~~~~~~
                \irule{}
                      {\Gamma', A, C'_{1} \vdash C'_{2}, E, \Delta'}
                      {\mbox{$(C')$ Axiom}}
               }
        {\Gamma', A \vdash C'_{2}, E, \Delta'}
        {\mbox{$(C'_{1})$~Cut}}
         ~~~~~~~~~~~~~~~~~~~~~~~~~~~~~~~~~~
         \irule{}
               {\Gamma', A, C'_{2} \vdash E, \Delta'}
               {\mbox{$(D)$ Axiom}}
        }
        {\Gamma', A \vdash E, \Delta'}
        {\mbox{$(C'_{2})$~Cut}}$}

\noindent otherwise this proof cannot be reduced.
\end{definition}

\begin{example}
This proof reduction algorithm may fail. 
Consider the rewrite system 
$$a \ra b~~~a \ra b'$$
There is no way to reduce the proof   
$$\irule{\irule{}
               {P(b) \vdash P(a), P(b')}
               {\mbox{$(P(b))$ Axiom}}
         ~~~~~~~~~~~~~~~~~~~~~~~~
        \irule{}
              {P(b), P(a) \vdash P(b')}
              {\mbox{$(P(b'))$ Axiom}}
        }
        {P(b) \vdash P(b')}
        {\mbox{$(P(a))$ Cut}}$$
\end{example}

But this situation cannot occur if the rewrite system is locally confluent.

\begin{proposition}
If the rewrite system is locally confluent, then the proof reduction
algorithm of definition \ref{asymmetricreduction} does not fail.
\end{proposition}

\proof{
If $C'_{1} \la^{1} C \ra^{1} C'_{2}$ then, by local confluence, there
is a proposition $C'$ such that $C'_{1} \ra^{*} C' \la^{*} C'_{2}$.}

\begin{example}
\label{loop}
The proof reduction algorithm of definition \ref{asymmetricreduction} may loop.
Consider the rewrite system 
$$a \ra b~~~a \ra c~~~b \ra a~~~b \ra d$$
Let $A$ be the proposition $P(a)$, $B$ be the
proposition $P(b)$,  $C$ be the proposition $P(c)$ and $D$ be the
proposition $P(d)$. 
We write $D^{n}$ for the proposition $D$ repeated $n$ times.
The proof 
$$\irule{\irule{}
               {C \vdash A, D, D^{n}}
               {\mbox{$(C)$ Axiom}}
         ~~~~~~~~~~~~~~~~~
         \irule{}
               {C, A \vdash D, D^{n}}
               {\mbox{$(D)$ Axiom}}
        }
        {C \vdash D, D^{n}}
        {\mbox{$(A)$ Cut}}$$
reduces to 
$$\irule{\irule{\irule{}
                      {C \vdash C, B, D, D^{n}}
                      {\mbox{$(C)$ Axiom}}
                ~~~~~~~~~~~~~~~~~
                \irule{}
                      {C, C \vdash B, D, D^{n}}
                      {\mbox{$(C)$ Axiom}}
               }
               {C 
\vdash B, D, D^{n}}
               {\mbox{$(C)$ Cut}}
         ~~~~~~~~~~~~~~~~~~~~~~~~~~~~~~~~~~
         \irule{}
               {C, B \vdash D, D^{n}}
               {\mbox{$(D)$ Axiom}}
        }
        {C \vdash D, D^{n}}
        {\mbox{$(B)$ Cut}}$$
that reduces to 
$$\irule{\irule{}
               {C \vdash B, D, D^{n}}
               {\mbox{$(C)$ Axiom}}
         ~~~~~~~~~~~~~~~~
         \irule{}
               {C, B \vdash D, D^{n}}
               {\mbox{$(D)$ Axiom}}
        }
        {C \vdash D, D^{n}}
        {\mbox{$(B)$ Cut}}$$
that reduces to 
$$\irule{\irule{\irule{}
                      {C \vdash A, D, D, D^{n}}
                      {\mbox{$(C)$ Axiom}}
                ~~~~~~~~~~~~~~~~
                \irule{}
                      {C, A \vdash D, D, D^{n}}
                      {\mbox{$(D)$ Axiom}}
               }
               {C \vdash D, D, D^{n}}
               {\mbox{$(A)$ Cut}}
         ~~~~~~~~~~~~~~~~~~~~~~~~~~~~~~~~~
         \irule{}
               {C, D \vdash D, D^{n}}
               {\mbox{$(D)$ Axiom}}
        }
        {C \vdash D, D^{n}}
        {\mbox{$(D)$ Cut}}$$
that contains the initial proof (for $n+1$) as a sub-proof. The proof
reduction algorithm loops on this proof, replacing a cut on
the proposition $A$ by one on the proposition $B$ and vice versa.
\end{example}

But this situation cannot occur if the rewrite system is 
terminating.

\begin{proposition}
If the rewrite system is terminating then the proof reduction
algorithm of definition \ref{asymmetricreduction} is terminating.
\end{proposition}

\proof{As the rewrite system is terminating, its reduction ordering is 
well-founded and thus so is the multiset extension of this ordering.

At each step, the algorithm of definition \ref{asymmetricreduction} 
replaces a cut with the cut proposition $C$ by two cuts with the cut 
propositions $C'_{1}$ and $C'_{2}$ where $C \ra^{1} C'_{1}$ and 
$C \ra^{1} C'_{2}$. Thus, the multiset of cut propositions in the proof 
decreases for the multiset extension of the reduction ordering of the 
rewrite system. Therefore, the proof reduction algorithm terminates.} 

\begin{corollary}
If a rewrite system is locally confluent and terminating then 
the proof reduction algorithm of definition \ref{asymmetricreduction}
always succeeds.
\end{corollary}

\begin{corollary}
If a rewrite system is locally confluent and terminating then 
asymmetric deduction modulo this rewrite system has the cut
elimination property.
\end{corollary}

\begin{corollary}[Newman's theorem \cite{Newman}]
If a rewrite system is locally confluent and terminating then 
it is confluent.  
\end{corollary}

\begin{remark}
We have seen that confluence is equivalent to cut elimination and that 
local confluence and termination imply normalization (i.e. termination
of the proof reduction algorithm) and hence cut elimination.
But notice that confluence alone does not imply normalization.
Indeed, if we add to the rewrite system of example \ref{loop} the
rules $c \ra e$ and $d \ra e$, we obtain confluence and thus cut
elimination, but the counterexample to normalization still holds. We
obtain this way an example of system that has the cut elimination
property, but not the normalization property.  Hence, in asymmetric
deduction modulo, normalization is a stronger property than cut
elimination.

In \cite{Newman}, Newman proves more than confluence (cut elimination)
for terminating locally confluent rewrite systems, as he proves
normalization, i.e. the termination of an algorithm that reduces peaks
step by step in conversion sequences.
\end{remark}

\section{Cut elimination in full asymmetric deduction modulo}

We consider now full asymmetric deduction modulo, and we prove 
that cut elimination is still equivalent to the confluence of the
rewrite system. 

\begin{proposition}
The cut rule is redundant in asymmetric deduction modulo a
rewrite system if and only if this rewrite system is confluent. 
\end{proposition}

\proof{The fact that cut elimination implies confluence is easy as 
cut elimination implies cut elimination for the atomic case and
hence confluence.

To prove that confluence implies cut elimination, we have to extend
the proof of proposition \ref{main} to the non atomic case. 

Without loss of generality, we can restrict to proofs where the 
axiom rule is used on atomic propositions only. 

We follow the cut elimination proof of \cite{Giraflor}. 
When we have a proof containing a cut 
$$\irule{\irule{\pi_{1}}{\Gamma, C_{1} \vdash \Delta}{}
         ~~~
         \irule{\pi_{2}}{\Gamma \vdash C_{2}, \Delta}{}
        }
        {\Gamma \vdash \Delta}
        {\mbox{$(C)$ Cut $C_{1} \la^{*} C \ra^{*} C_{2}$}}$$
then we show
that from $\pi_{1}$ and $\pi_{2}$, we can reconstruct a proof of
$\Gamma \vdash \Delta$ introducing cuts on smaller propositions than $C$
(i.e. propositions with fewer connectors and quantifiers).

More generally, we prove, by induction on the structure of $\pi_{1}$
and $\pi_{2}$, that from a proof $\pi_{1}$ of $\Gamma,
\overline{C}_{1}
\vdash \Delta$ and $\pi_{2}$ of  $\Gamma \vdash \overline{C}_{2},
\Delta$, where $\overline{C}_{1}$ and $\overline{C}_{2}$ are multisets
of reducts of some proposition $C$, we can reconstruct a proof $\Gamma
\vdash \Delta$. 
Notice that, as the rewrite system rewrites terms only, rewriting does
not change the logical structure of a proposition (i.e. an atomic
proposition only rewrites to an atomic proposition, a conjunction
to a conjunction, ...)

There are several cases to consider.
\begin{itemize}
\item If the last rule of $\pi_{1}$ or the last rule of $\pi_{2}$ is 
a structural rule, then we apply the induction hypothesis.

\item If the last rule of $\pi_{1}$ or the last rule of $\pi_{2}$ is 
a logical rule on a proposition in $\Gamma$ or $\Delta$, then 
we apply the induction hypothesis.

\item If the last rule of $\pi_{1}$ or the last rule of $\pi_{2}$ is 
an axiom rule on propositions in $\Gamma$ and $\Delta$, then
$\Gamma$ and $\Delta$ contain two propositions that have a common
reduct $C$
and we take the proof 
$$\irule{}{\Gamma \vdash \Delta}{\mbox{$(C)$ Axiom}}$$

\item
The key case in the proof of \cite{Giraflor} 
is when both $\pi_{1}$ and $\pi_{2}$ end with a logical rule on a
proposition in $\overline{C}_{1}$ and $\overline{C}_{2}$.
For instance, if $C$ has the form $A \wedge B$, 
$A_{1} \wedge B_{1} \la^{*} A \wedge B \ra^{*} A_{2} \wedge B_{2}$
and 
the proofs $\pi_{1}$ and $\pi_{2}$ have the form
$$\irule{\irule{\rho_{1}}
               {\Gamma, A'_{1}, B'_{1} \vdash \Delta}
               {}
        }
        {\Gamma, A_{1} \wedge B_{1} \vdash \Delta}
        {\mbox{$\wedge$-left}}$$
with 
$A_{1} \wedge B_{1} \ra^{*} 
A'_{1} \wedge B'_{1}$
and
$$\irule{\irule{\rho_{2}}
               {\Gamma \vdash A'_{2}, \Delta}
               {}
         ~~~~~
         \irule{\rho_{3}}
               {\Gamma \vdash B'_{2}, \Delta}
               {}
        }
        {\Gamma \vdash A_{2} \wedge B_{2}, \Delta}
        {\mbox{$\wedge$-right}}$$
with $A_{2} \wedge B_{2} \ra^{*} A'_{2} \wedge B'_{2}$.

In this case, we have
$A'_{1} \wedge B'_{1} \la^{*} 
A_{1} \wedge B_{1} \la^{*} 
A \wedge B 
\ra^{*} A_{2} \wedge B_{2}
\ra^{*} A'_{2} \wedge B'_{2}$, thus 
$A'_{1} \la^{*} A \ra^{*} A'_{2}$
and $B'_{1} \la^{*} B \ra^{*} B'_{2}$
and we reconstruct the proof
$$\irule{\irule{\irule{\rho_{1}}
                      {\Gamma, B'_{1}, A'_{1} \vdash \Delta}
                      {}
                ~~~
                \irule{\irule{\rho_{2}}
                             {\Gamma \vdash A'_{2}, \Delta}
                             {}
                      }
                      {\Gamma, B'_{1} \vdash A'_{2}, \Delta}
                      {\mbox{weak-left}}
               } 
               {\Gamma, B'_{1} \vdash \Delta}
        {\mbox{$(A)$ Cut}}
         ~~~~~~~~~~~~~~~~~~~~~~~~~~~~~~
         \irule{\rho_{3}} 
               {\Gamma \vdash B'_{2}, \Delta}
               {}
        }
        {\Gamma \vdash \Delta}
        {\mbox{$(B)$ Cut}}$$
The case of the other connectors and quantifiers is similar.

\item The new case in asymmetric deduction modulo is 
when the last rule of both proofs is an axiom rule involving a
proposition in $\overline{C}_{1}$ and $\overline{C}_{2}$. Notice that
the proposition $C$ is atomic in this case, as we have restricted 
the Axiom rule to apply on atomic propositions only.
Thus, $\Gamma$ contains a
proposition $A$ that has a common reduct $B$ with $C_{1}$ in
$\overline{C}_{1}$ 
and $\Delta$
contains a proposition $E$ that has a common reduct $D$ with
$C_{2}$ in $\overline{C}_{2}$. 
We have 
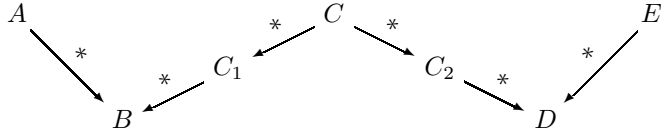
\begin{diagram}[height=1em,width=2em]
A&&&&&&C&&&&&&E\\
&\rdTo(2,4)^*&&&&\ldTo^*&&\rdTo^*&&&&\ldTo(2,4)^*&\\
&&&&C_1&&&&C_2&&&&\\
&&&\ldTo^*&&&&&&\rdTo^*&&&\\
&&B&&&&&&&&D&&\\
\end{diagram}
This is the case where we use confluence to obtain that $A$ and $E$
have a common reduct $C'$ and we take the proof 
$$\irule{}{\Gamma \vdash \Delta}{\mbox{$(C')$ Axiom}}$$
\end{itemize}}

\begin{remark}
This proof suggests a cut elimination algorithm that integrates the
cut elimination algorithm of sequent calculus and Newman's
algorithm: it behaves like the latter for atomic cuts and like the
former for non atomic ones. Again, we need the termination of the rewrite 
system to prove the termination of this cut elimination algorithm.
\end{remark}

\begin{remark}
If we consider now rules directly rewriting atomic propositions to non atomic
ones: for instance a rule like $A \ra B \wedge \neg A$ then confluence
\cite{Crabbe74,DowekWerner-normalization-98}, and even confluence and
termination \cite{DowekWerner00}, do not imply cut elimination
anymore. Some propositions have proofs using the cut rule but no cut
free proof. 

In this case, confluence is not a sufficient analyticity condition
anymore. As a consequence, with term rewrite systems, confluence is a
sufficient condition for the completeness of proof search methods,
such as equational resolution (i.e. resolution where some equational
axioms are dropped and unification is replaced by equational
unification), but with proposition rewrite systems, confluence is not
a sufficient condition for the completeness of resolution modulo and
this condition must be replaced by cut elimination (see
\cite{DowekFrocos} for a discussion on this point).
\end{remark}

\section*{Conclusion}

When a congruence is defined by a term rewrite system, the
confluence of this rewrite system and the cut elimination property for
asymmetric deduction modulo this system coincide and analyticity can
be defined either using one property of the other.

When the rewrite system is also terminating, then asymmetric deduction
modulo not only verifies cut elimination, but also normalization.

But when a congruence is defined by rules directly rewriting 
atomic propositions, confluence is not a sufficient analyticity 
condition anymore and must be replaced by cut elimination.

\section*{Acknowledgments}
I want to thank Th\'er\`ese Hardin, Claude Kirchner and Benjamin
Werner for comments on a draft of this paper and for many discussions
about deduction modulo, confluence, termination and cut elimination.

\end{document}